\documentclass[a4paper,english]{iaus}
\usepackage[T1]{fontenc}
\usepackage[latin9]{inputenc}
\usepackage{graphicx}

\makeatletter

\newcommand{\noun}[1]{\textsc{#1}}

\newcommand{\kms}{\,km~s$^{-1}$}

\def\arcmin{\hbox{$^\prime$}}
\def\arcsec{\hbox{$^{\prime\prime}$}}

\makeatother

\usepackage{babel}

\begin{document}
\title[NGC 205]{Morphological transformation of NGC 205?} 
\author[Saviane, Monaco \& Hallas]{Ivo Saviane$^1$, Lorenzo Monaco$^2$ \and Tony Hallas$^3$}
\affiliation{$^1$European Southern Observatory,  Chile \\ email: {\tt isaviane@eso.org} \\[\affilskip] $^2$Dept. of Astronomy, University of Concepcion,  Chile  \\[\affilskip] $^3$Astrophoto,  USA } 
\pubyear{2009} \volume{262}
\pagerange{xxx--xxx} \setcounter{page}{1} \jname{Stellar Populations - Planning for the Next Decade}
\editors{G. Bruzual \& S. Charlot, eds.} 
\maketitle

\firstsection % if your document starts with a section,

\noindent \vspace*{-5mm}

\section*{A dE galaxy with recent central star formation}

\noindent NGC~205 is a small galaxy ($M/M_{\odot}=0.7\times10^{9}$;
$M_{V}=-16.6$) currently located $36\arcmin$ NW of M31. It is classified
as dE because in ground-based images it appears as an elliptical body.
However past investigations have revealed characteristics that are
more typical of a disk galaxy: the specific frequency of globular
clusters is $1.8$; the large scale dynamics shows partial rotational
support; there is a significant amount ($10^{6}\, M_{\odot}$) of
rotating gas (molecular and atomic) and dust; the central regions
harbor a fairly complex stellar population, including a $100$--$500$~Myr
old nucleus surrounded by $50$- and $100$-Myr old stellar associations
(see references in Monaco et al. \cite{monaco_etal09}; M09). Very
recently, thanks to \noun{hst/acs} imaging we have been able to reveal
a young central `field' population (M09), extending out to $\sim40\arcsec$
in radius ($\sim160$~pc). The luminosity function of the main sequence
can be fitted with Saviane et al. (\cite{saviane_etal04}) model of
continuous star formation (SF) from at least $\sim600$~Myr ago to
$\sim60$~Myr ago. We found that  $1.5\times10^{5}\, M_{\odot}$
in stars were produced from $\sim300$~Myr to $\sim60$~Myr ago,
with a SF rate of $7\times10^{-4}\, M_{\odot}$~yr$^{-1}$. A continuous
SF seems to support the latest simulations of NGC~205 orbit: Howley
et al. (\cite{howley_etal08}) found that the galaxy must be moving
with a velocity $300-500$~\kms (comparable to the escape velocity)
along an almost radial orbit, and it should be approaching M31 for
the first time. An episodic SF triggered by passages through M31 disk
every $\sim300$~Myr in a bound orbit (Cepa \& Beckman \cite{cepa_beckman88})
is excluded by our data. 

\noindent \vspace*{-5mm}

\section*{Substructures in and near the body of NGC 205}

\begin{figure}
\begin{centering}
\includegraphics[width=0.28\textwidth]{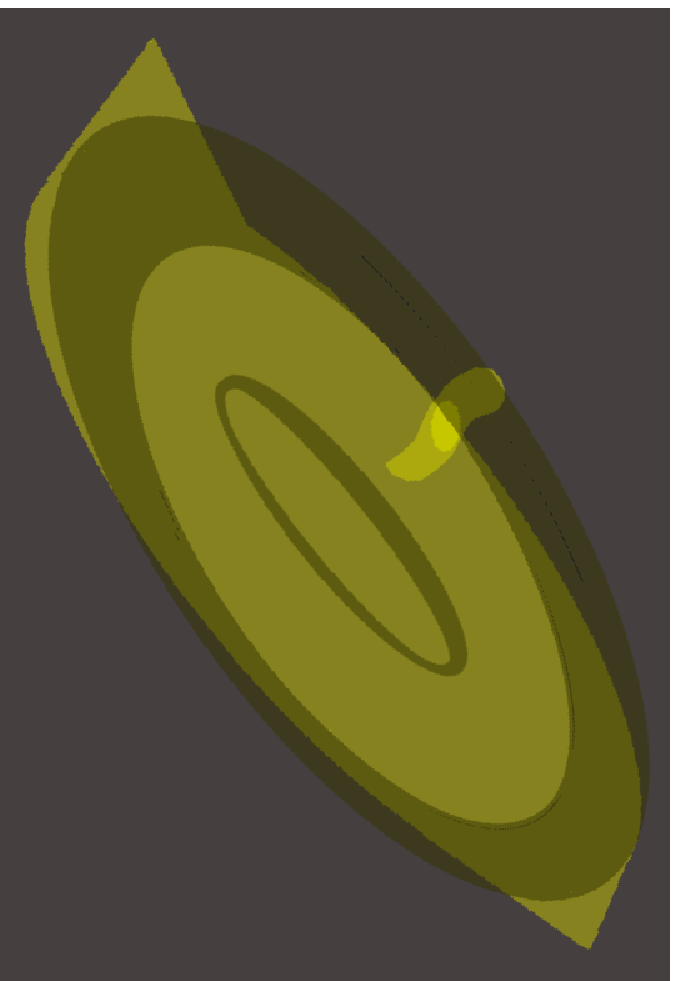}~\includegraphics[width=0.29\textwidth]{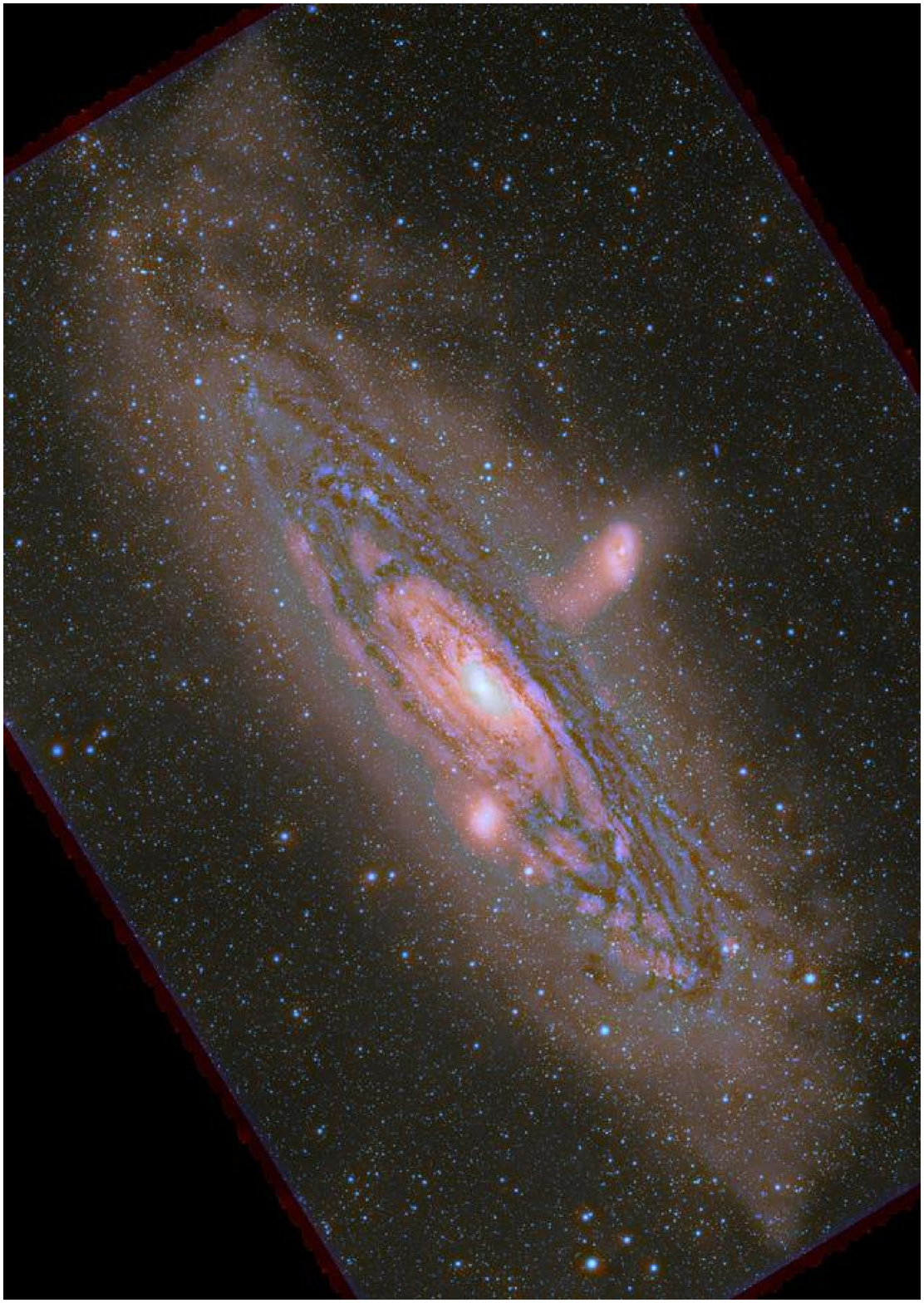}~\includegraphics[width=0.14\textwidth]{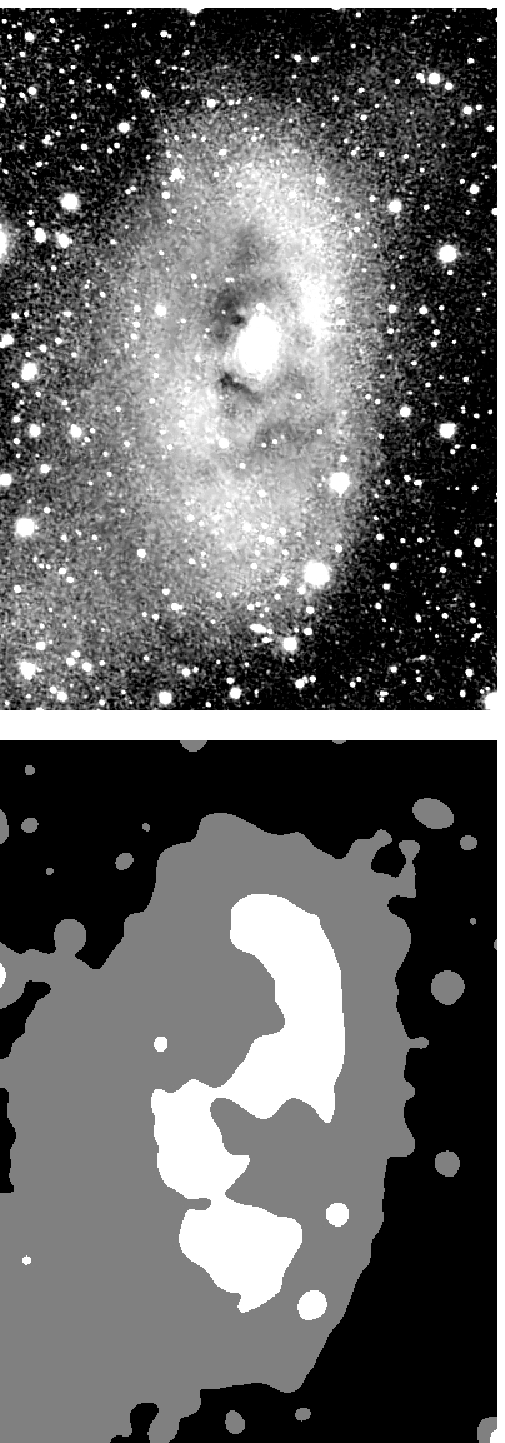}
\par\end{centering}

\caption{The central image of M31 and its satellites is based on a $9$-frame
mosaic taken with a SBIG/STL-11K camera and no filter, attached to
a $6$-inch f/8 Stellarvue refractor. The red channel has been obtained
by Gaussian smoothing with an 80~px window, while the blue channel
is the result of a four-pass unsharp-masking with a 50~px window.
The green channel is the original frame. The left panel is a schematic
representation of the distribution of stars and dust in M31, screening
NGC~205 and its tidal tails. The right panels show the unsharp masked
image of NGC~205 (top) and a three-level contour version of the same
frame (bottom). \label{fig:This-panoramic-and}}

\end{figure}

\noindent Kormendy and collaborators have proposed that all dE galaxies
were originally dwarf disks (e.g., Kormendy et al. \cite{kormendy_etal09}),
so the results discussed above can be placed in the broader context
of dE formation. In particular, Lisker et al. (\cite{lisker_etal06b})
discovered a number of blue-center dEs in the Virgo cluster, and considered
NGC~205 a local representative of this new dE(bc) class (see also
Koleva et al. \cite{koleva_etal09}). In Virgo, dE(bc) galaxies present
flattening distributions indicating disk shapes and show no central
clustering, which leads the authors to suggest galaxy harassment (Moore
et al. \cite{moo96}) as their likely formation mechanism. Harassment
requires repeated encounters with large galaxies, so it should not
work in a low-density group environment. It is therefore suggested
that `tidal stirring' (Mayer et al. \cite{mayer_etal01}) could lead
to morphological transformation in the case of NGC~205. An alternative
mechanism is that of `galaxy threshing' (Bekki et al. \cite{bekki_etal01}).
In the course of transformation gas is funneled to the center and
forms a density excess which could explain the blue centers (Moore
et al. \cite{moo98}; Bekki et al. \cite{bekki_etal01}; Mayer et
al. \cite{mayer_etal01}). The rise of SF in the center of NGC~205
in correspondence to a decline in the external regions (Koleva \cite{koleva09})
might be a signature of this phenomenon. If the dE is at its first
encounter with M31, we might be seeing it at the earliest phases of
this process, when still retaining most of its disk features. In the
harassment scenario, bar and spiral features can be created and retained
for some time. Indeed Lisker et al. (\cite{lisker_etal06b}) present
the case of VCC~0135 and VCC~1437 as showing possible spiral arms
and a bar. In Fig.~\ref{fig:This-panoramic-and} we show that some
sub-structure is also present inside the body of NGC~205, thanks
to a two hour exposure frame, at a scale of $\sim2\arcsec$~px$^{-1}$.
The southern tidal tail stands out in a very prominent way, out to
at least $17\arcmin$ ($\sim3.5$~kpc) from the galaxy center. The
tail is probably longer, but it is hidden behind the high-extinction
central regions of M31. The near absence of a northern tail might
be explained by a very extended M31 dust ring screening it. This is
plausible because NGC~205 is farther than M31 and the near-side of
M31 is the NW one (see the schematic sketch in Fig.~\ref{fig:This-panoramic-and}).
Figure~\ref{fig:This-panoramic-and} further shows that the well-known
dust clumps of NGC~205 seem to be part of a ring surrounding the
central nucleus. Moreover, sub-structure is seen also across the main
body of the galaxy, which resembles spiral arms connected by a bar.
While this feature is quite intriguing in the context of morphological
transformation, a foreground dust filament might also be responsible
for the illusion. Very deep and wide-field imaging in three bands,
such as that offered by the \noun{hst/acs}, would be needed to discriminate
between the two hypotheses. 

\end{document}